\documentclass[prb,aps,reprint,amsmath,amssymb,twocolumn,floatfix]{revtex4-1}
\usepackage{graphicx}% Include figure files
\usepackage{epsfig} % Include figure files
\usepackage{textcomp}% for weird symbols
\usepackage{natbib}% fancy ref

\begin{document}

\title{A neural network interatomic potential for the phase change material GeTe}

\author{Gabriele C. Sosso$^{1}$}
\author{Giacomo Miceli$^{1}$}
\altaffiliation{Current address: Institute of Theoretical Physics, \'{E}cole Polytechnique F\'{e}d\'{e}rale de Lausanne (EPFL), 
CH-1015 Lausanne, Switzerland}
\author{Sebastiano Caravati$^{2}$}
\author{J\"{o}rg Behler$^{3}$}
\author{Marco Bernasconi$^{1}$}\email[Corresponding author. Email address: ]{marco.bernasconi@unimib.it}

\affiliation{
$^1$ Dipartimento di Scienza dei Materiali, Universit\`{a} di Milano-Bicocca,
Via R. Cozzi 53, I-20125 Milano, Italy}

\affiliation{
$^2$ Computational Science, Department of Chemistry and Applied Biosciences,
ETH Zurich, USI Campus, via Giuseppe Buffi 13, CH-6900 Lugano,
Switzerland}

\affiliation{
$^3$ Lehrstuhl f\"{u}r Theoretische Chemie, Ruhr-Universit\"{a}t Bochum, D-44780 Bochum, Germany
}

\date{\today}

\begin{abstract}
GeTe is a prototypical phase change material
of high interest for applications in optical and electronic non-volatile memories.
We present an interatomic potential for the bulk phases of GeTe, which
is created using a neural network (NN) representation of the potential-energy surface obtained from
reference calculations based on density functional theory. 
It is demonstrated that the NN potential provides a close to ab initio quality description of a number of properties
of liquid, crystalline and amorphous GeTe. The availability of a reliable classical potential allows
addressing a number of issues of interest for the
technological applications of phase change materials, which are presently beyond the capability of first 
principles molecular dynamics simulations.
\end{abstract}


\maketitle

\section{\label{introduction} INTRODUCTION}
Phase-change materials based on chalcogenide alloys are attracting an increasing interest worldwide
due to their ability to undergo reversible
and fast transitions between the amorphous and crystalline phases upon
heating~\cite{1}. This property is exploited in rewriteable optical media (CD, DVD, Blu-Ray Discs) and electronic
nonvolatile memories (NVM), which are based on the strong optical and electronic contrast between the two phases~\cite{2,3}.
The material of choice for NVM applications is the ternary compound Ge$_2$Sb$_2$Te$_5$ (GST). However, the related binary alloy
GeTe has also been thoroughly investigated because of its higher crystallization temperature and better data retention at
high temperature with respect to GST.

In the last few years atomistic simulations based on density functional theory (DFT) have provided useful insights into the properties of
phase change materials~\cite{lencerReview,welrev,cara07,cara09,akola}. However, several key issues such as the thermal 
conductivity at the nanoscale, the crystallization dynamics, and
the properties of the crystalline/amorphous interface, just to name a few, are presently beyond the reach of  
ab initio simulations due to the high computational costs. The development of reliable classical interatomic potentials is a possible route to overcome
the limitations in system size and time scale of ab initio molecular dynamics.
However, traditional approaches based on the fitting of comparably simple functional forms for the
interatomic potentials are very challenging due to the complexity of the
chemical bonding in the crystalline and amorphous phases revealed by the ab initio simulations. A
possible solution has been proposed recently by Behler and Parrinello~\cite{7}, who developed high-dimensional
interatomic potentials with close to ab initio accuracy employing artificial neural networks (NN). To date, potentials
of this type have been reported for silicon~\cite{8,pssbbehler}, carbon~\cite{9,khaliullin2011}, sodium~\cite{10}, zinc oxide~\cite{11} and 
copper~\cite{artrith} by fitting large ab initio databases.

Here we describe the development of a classical interatomic potential for the bulk phases of GeTe employing this NN technique.
The potential is validated by comparing results on the structural and dynamical properties of liquid,
amorphous and crystalline GeTe derived from NN-based simulations with the ab initio data obtained in our previous work~\cite{mazza}.

\section{\label{methods} METHODS}

\subsection{\label{coming_closer}  The Neural Network Method}

\begin{figure*}[htbp!] 
\centerline{\includegraphics[width=1.6\columnwidth]{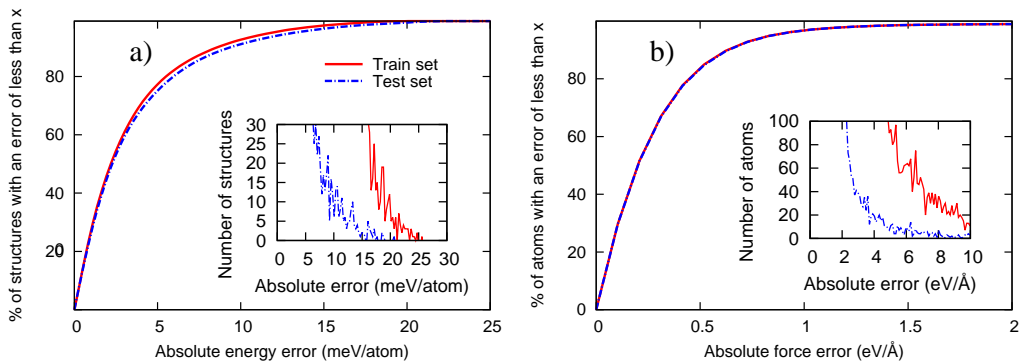}}
\caption{Normalized cumulative histograms of the absolute NN errors in training and test data sets for the energies (a) and forces (b).
Standard histograms for the same data are presented as insets.}
\label{FIG1}
\end{figure*}

Artificial neural networks constitute a class of algorithms inspired by the properties of biological
neural networks, which are widely applied in many different fields
ranging from weather forecasting to robotics~\cite{bishop}. In the last few years,
NN have also been used as a tool to construct interatomic potentials~\cite{handely,pccpbehler}
To this aim, the NN are exploited as a non-linear technique that allows fitting any real-valued function to arbitrary accuracy,
without any previous knowledge about the functional form of the underlying problem~\cite{hornik,cybenko}.
In our case, this function is the potential energy surface (PES) of the atomistic system and the goal
of the NN is to construct a functional relation between the energy and the atomic configuration.

Given a number of atomic configurations, for which the ab initio energies are known, the NN parameters are adjusted to
reproduce these energies in the training process. 
Overfitting, i.e., obtaining a good fit of the training data, but performing less accurately when
making predictions is controlled by testing the performance of the NN for an independent test set not used in the fitting.
Once trained, the NN performs an interpolation to construct the potential energy for new atomic configurations
at a low computational load, which allows performing long molecular dynamics runs for large systems.

The NN methodology overcomes many problems associated with classical potentials. First and foremost,
NN completely obviate the problem of guessing a complicated functional form of the interatomic potential. This form is
determined automatically by the NN. Moreover, the entire training procedure is fully automated so that NN can be
readily extended to new regions of the PES. Finally, the accurate mapping of ab initio energies ensures
that all properties determined by the topology of the
PES are described with an accuracy comparable with that of first principles calculations.
NN have been successfully used to interpolate the PES of simple chemical systems like small 
molecules~\cite{gassner,p0420,p0831,p0826,p2391} 
and molecule-surface interactions~\cite{blank,lorenz03,p1786,behler2007,behler2008,carbogno}.

Until recently, the main limitation of most NN approaches has been the small number of degrees of freedom
that could be described, which confined their application to very small systems. 
This limitation has been overcome by Behler and Parrinello who introduced a NN
potential, in which the total energy is expressed as a sum of atomic energy contributions depending on the local environments
\cite{7}. In this approach the energy contribution of each atom is evaluated using an 
individual NN instead of using one NN for the total energy of the system.
The local environment of a given atom is described by a set of local parameters called symmetry functions, which include
radial and angular many-body terms and depend on the positions of all neighbors within a specified cutoff radius~\cite{17}.
The use of symmetry functions instead of Cartesian coordinates as NN inputs and the partitioning of the total energy
into atomic contributions ensures that all quantities computed with the NN potential, such as energies, analytical forces, and stress
tensor are invariant to translations, rotations, and atoms exchange. Furthermore, once the fit is obtained, the NN
potential can be applied to systems containing an arbitrary number of atoms.
The validity of this approach has been demonstrated by reproducing
the high-pressure phase diagram of silicon, carbon, and sodium~\cite{8,9,10}.
This scheme has been recently extended to binary systems by including
long range Coulomb interaction between environment dependent ionic charges~\cite{11,morawietz}.

In the case of GeTe, we are faced with the problem of developing a potential suitable
to describe both the semiconducting crystalline and amorphous phases as well as the metallic liquid.
As a first step toward the development of a NN potential for GeTe, we 
neglect long-range Coulomb interactions for atoms being separated by a larger distance than the cutoff radius of the
symmetry functions. The resulting ``short-ranged'' NN potential just consists of atomic energy contributions arising from
the local chemical environments, but it is important to note that also short-ranged electrostatic interactions are 
fully taken into account implicitly. More details of the employed method can be found elsewhere~\cite{7,17}.
The resulting potential will not be able to describe the dielectric response and LO-TO splitting of crystalline GeTe
in its ferroelectric phase, but it will be suitable to
reproduce structural properties of the liquid, amorphous and crystalline phases and the dynamical properties of the disordered
phases.

\subsection{\label{a_nn} The NN potential for GeTe}

To generate the NN potential, we fitted the  total  energies   of about 30.000
configurations of 64-, 96- and 216-atom supercells computed by DFT.
We started with a relatively small dataset of about 5.000 structures, where
we considered crystalline configurations, snapshots of the liquid phase and of
the amorphous phase generated by quenching from the melt at ambient conditions and at 
different pressures up to 50 GPa. 
We also considered mixed crystalline/amorphous models generated by partially crystallizing the amorphous phase by means
of the metadynamics technique~\cite{15}. All these configurations were generated within ab initio molecular dynamics
simulations at different temperatures (up to 3000 K) with the code CP2K~\cite{16}.
This first dataset was then  expanded by adding randomly distorted structures of the initial dataset at slightly different
pressures and temperatures and models with slight deviations from the perfect stoichiometry.
The refinement of the potential was achieved
by inserting in the data set the ab initio energy of configurations generated by Molecular Dynamics simulations (see below) 
using the not yet refined NN potential.

Total energy convergence has to be strictly guaranteed for every point of the dataset to make the NN input fully
consistent. Therefore, we had to perform Brillouin Zone (BZ) integration over a dense $4\times4\times4$ Monkhorst-Pack (MP)~\cite{MP}
mesh for the 64-atom cell and employing meshes of a corresponding k-point density for the larger systems.
The ab initio energies were calculated using the QUANTUM-ESPRESSO
package~\cite{12}. The Perdew-Burke-Ernzerhof (PBE)~\cite{13} exchange-correlation functional and norm
conserving pseudopotentials were employed, considering only the outermost $s$ and $p$ electrons in the valence shell.
The Kohn-Sham  orbitals were expanded in a plane waves basis up to a kinetic energy cutoff of 40~Ry.
These settings ensure convergence of the total energy to 2~meV/atom.

The best NN fit we found employs three hidden layers with 20 nodes each. Sigmoidal activation functions were used in the 
nodes of the hidden layers, while a linear function was used for the output node.
Details concerning the architecture of the NN can be found elsewhere~\cite{17}.
The local environment of each atom is defined by the value of 159 symmetry functions (see Ref.~\cite{17} for details)
defined in terms of the positions of all neighbors within a distance cutoff of 6.88~\AA.
The generation of the NN potential and the calculation of the forces for the MD simulations
were performed with the NN code RuNNer~\cite{21}.
We used the TINKER~\cite{22} code as MD driver. The time step for the MD runs was set to 0.2~fs, and constant temperature simulations
were performed using the Berendsen thermostat~\cite{bere}.

The results of the fitting process of the NN potential are summarized in Fig.~\ref{FIG1}. The root mean square error (RMSE) for the energy is
5.01 and 5.60 meV/atom for the training and the test set, respectively, while the RMSEs of the forces are 0.46 and 0.47 eV/\AA,
for the two sets. Among all structures considered, only a negligible fraction
shows noticeable absolute errors, up to 25.8 meV/atom and 11.2 eV/\AA$\ $ for energies 
and forces (see Fig. \ref{FIG1} insets). We have found that these configurations correspond to high-energy structures that are not
visited in MD simulations carried out in the present work.

\section{\label{results} RESULTS}

\subsection{\label{crystal} Crystalline phase}

\begin{figure}[htbp!]
\centerline{\includegraphics[width=0.5\columnwidth]{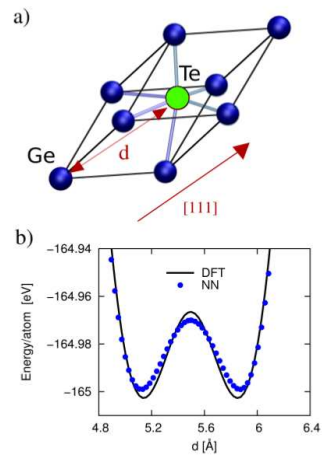}}
\caption{a) Trigonal structure of crystalline GeTe, which can be seen as a cubic rocksalt geometry with an elongation of
the [111] diagonal. Ge atoms are on the vertexes of the cell while Te atom stands in the center. 
The distance between Ge and Te atoms along the [111] direction is indicated by $d$. b) Energy of  trigonal GeTe
 as a function of $d$ at fixed values of the lattice parameters at the theoretical equilibrium geometry (cf. Table \protect\ref{TAB1}).}
\label{FIG2}
\end{figure}

The equilibrium geometry of the trigonal phase of crystalline GeTe  ($R{3}m$ space group) was obtained
by optimizing all structural parameters consisting of the lattice parameter $a$, the trigonal angle $\alpha$ and the internal
parameter $x$ that assigns the positions of the two atoms in the unit cell, namely Ge  at ($x$,$x$,$x$) and  Te at (-$x$,-$x$,-$x$)~\cite{goldak}. 
The residual anisotropy in the stress tensor at the optimized lattice parameter at each volume is below
0.02~kbar.  The energy versus volume data were fitted with a Murnaghan equation of state~\cite{murna}.
The theoretical structural parameters of the trigonal phase of GeTe at equilibrium are compared in Tab.~\ref{TAB1} 
with experimental data~\cite{goldak} and DFT results obtained with a $12\times12\times12$ MP k-point mesh in the BZ integration.
DFT data are similar to those reported previously~\cite{lencer}. The length of the short and long Ge-Te bonds are also given.
The structure of trigonal GeTe can be seen as a distorted rocksalt geometry with an elongation of the cube diagonal
along the [111] direction and an off-center displacement of the inner Te atom along the [111] direction, which moves to a distance
$d$ from the Ge atom at the vertex as shown in Fig.~\ref{FIG2}a. The energy gained by the off-center displacement is analyzed
by varying the distance $d$  at fixed lattice parameters $a$ and $\alpha$ .
The resulting energy as a function of $d$ is reported in  Fig. \ref{FIG2}b for the NN and the DFT calculations.
We note that the DFT values were not included in the training set but were recalculated for investigating the
quality of the NN potential only.
The double well potential identifies the two possible ferroelectric configurations while the maximum corresponds to a paraelectric
configuration.

\begin{table}[htbp]
\begin{center}
\begin{tabular}{c|c|c|c}
\hline
\hline
Structural parameters & NN & DFT  & Exp.\\
\hline
\hline
a (\AA)   & 4.47 & 4.33 & 4.31 \\
$\alpha$  & 55.07$^{\circ}$ & 58.14$^{\circ}$ & 57.9$^{\circ}$ \\
Volume (\AA$^3$) & 55.95 & 54.98 & 53.88 \\
x       &   0.2324      &  0.2358     &  0.2366   \\
Short, long bonds (\AA) & 2.81, 3.31 & 2.85, 3.21 & 2.84, 3.17 \\
\hline
\hline
\end{tabular}
\end{center}
\caption{Structural parameters of the trigonal phase of crystalline GeTe from  NN  and DFT  calculations
and from the experimental data of Ref. \citenum{goldak}. The lengths of the short and long bonds are also given.}
\label{TAB1}
\end{table}

As a further validation of the potential, we computed the difference in energy between the trigonal phase and an ideal rocksalt 
phase at their equilibrium volumes at zero temperature that amounts to 44 meV/atom or 55 meV/atom in NN and DFT calculations, respectively.

\begin{center}
\begin{table}[htbp]
\begin{tabular}{p{1.0cm}|p{0.5cm}|p{0.5cm}|p{0.5cm}|p{0.5cm}|p{0.5cm}|p{0.5cm}}
\hline
\hline
 & c$_{11}$ & c$_{12}$ & c$_{13}$ & c$_{14}$ & c$_{33}$ & c$_{44}$ \\
\hline
\hline
\hline
DFT & 92 & 18 & 22 & 35 & 40 & 24 \\
\hline
NN  & 73 & 10 & 30 & 24 & 36 & 20 \\ 
\hline
\hline
\end{tabular}
\caption{Elastic constants (GPa) of trigonal GeTe from DFT and NN calculations.}
\label{TAB2}
\end{table}
\end{center}

The elastic properties of trigonal GeTe were investigated by computing the elastic constants from finite deformations of the
lattice parameters. The NN and DFT results are compared in Table~\ref{TAB2}. The elastic constants obtained here with the PBE functional
are somehow softer than those obtained with the LDA functional in Ref.~\onlinecite{ratyPhon}. The bulk modulus obtained either from the elastic
constants or from the equation of state is 34 GPa and 33 GPa for the NN and DFT calculations.

\subsection{\label{liquid} Liquid phase}

\begin{figure}[htbp!]
\centerline{\includegraphics[width=0.7\columnwidth]{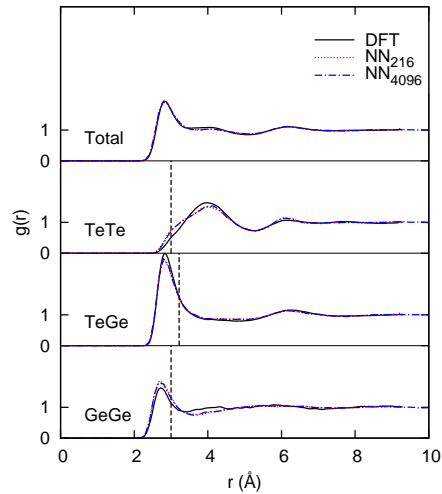}}
\caption{Total and partial pair correlation functions of liquid GeTe from a NN molecular dynamics simulation at 1150 K with 
a 4096-atom and a 216-atom cell, compared with results from an ab initio simulation at the same temperature using a 
216-atom cell \cite{mazza}. NN results are obtained by averaging over a NVE run 40 ps long at the average temperature of 1150 K.
 The vertical lines are the interatomic distance threshold used to define the coordination numbers, 3.0 \AA,$\, $ 3.22 \AA,$\, $ and 3.0 \AA $\, $ 
for Ge-Ge, Ge-Te and Te-Te bonds, respectively.}
\label{FIG4}
\end{figure}

\begin{figure}[htbp!]
\centerline{\includegraphics[width=0.5\columnwidth]{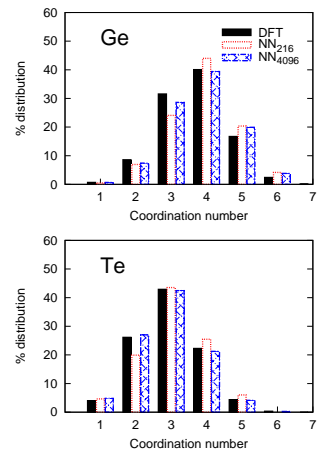}}
\caption{Distributions of coordination numbers of Ge and Te atoms in liquid GeTe at 1150 K. Results from NN 
(4096-atom and 216-atom) and ab initio (216-atom) \cite{mazza} simulations are compared.}
\label{coord}
\end{figure}

\begin{figure}[htbp!]
\centerline{\includegraphics[width=0.5\columnwidth]{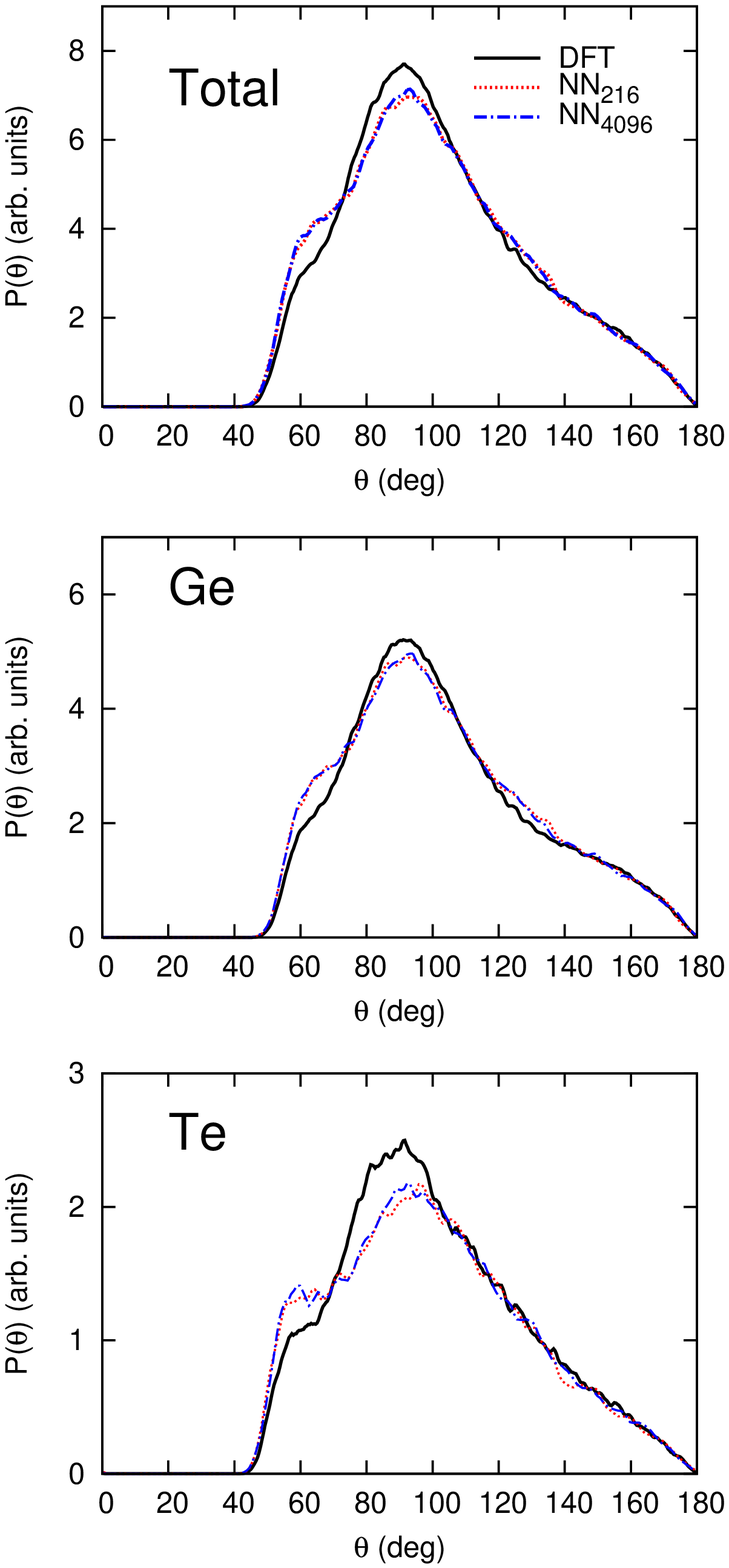}}
\caption{Total and partial angle distribution functions of liquid GeTe from a NN molecular dynamics simulation at 
1150 K with a 4096-atom and a 216-atom cell, compared with results from an ab initio simulation at the same 
temperature using a 216-atom cell \cite{mazza}. Partial distributions refer to X-Ge-Y and X-Te-Y triplets (X,Y= Ge or Te).}
\label{FIG5}
\end{figure}

The liquid phase of GeTe was simulated by a 4096-atom model at 1150 K.
Total and partial pair correlation functions are compared in Fig.~\ref{FIG4} with results from our previous ab initio 
simulations in a small 216-atom cell at the same temperature~\cite{mazza}. Results from the NN simulations of a 216-atom 
cell are also reported. The density of 0.03156 atoms/\AA$^3$ is the same for all simulations and corresponds
to the value chosen in the ab initio simulations of Ref.~\citenum{mazza}, which is close to the
experimental density of the amorphous phase~\cite{adens}.
Distributions of coordination numbers are reported in Fig.~\ref{coord} as computed by integrating the partial pair 
correlation functions up to the cutoff shown in Fig.~\ref{FIG4}. Average coordination numbers are given in Table~\ref{TAB3} while
 angle distribution functions are shown in Fig.~\ref{FIG5}. The agreement between NN and ab initio data is excellent. 
The NN results obtained with 216-atom and 4096-atom cells are extremely similar,
which demonstrates that structural properties of the liquid can be reliably described by the cells few hundred atoms large 
we used in our previous ab initio works~\cite{mazza,cara07,cara09,caraSbTe,spreafico}.
The self-diffusion coefficients computed from NN simulations are also in good agreement with the ab initio results of Ref.~\citenum{akola}
as shown in Table~\ref{TAB4}. These latter data refer to simulations at 1000 K to enable a comparison with previous ab initio
results obtained at this temperature.

\begin{table}[htbp]
\begin{center}
\resizebox{8cm}{!} {
\begin{tabular}{c|c|c|c|c|c|c|c|c|c}
\hline
\hline
 & \multicolumn{3}{c|}{With Ge} & \multicolumn{3}{|c|}{With Te} & \multicolumn{3}{|c}{Total} \\
\hline
 & DFT & NN$_{216}$ & NN$_{4096}$ & DFT & NN$_{216}$ & NN$_{4096}$ & DFT & NN$_{216}$ & NN$_{4096}$ \\
\hline
Ge & 1.00 & 1.11 & 1.15 & 2.71 & 2.78 & 2.67 & 3.71 & 3.89 & 3.82 \\
\hline
Te & 2.71 & 2.78 & 2.67 & 0.26 & 0.28 & 0.26 & 2.97 & 3.07 & 2.93 \\
\hline
\hline
\end{tabular}}
\end{center}
\caption{Average coordination numbers for different pairs of atoms computed from the partial pair correlation functions of 
liquid GeTe from a NN molecular dynamics simulation at 1150 K with a 4096-atom and a 216-atom cell (cf. Fig. \protect\ref{FIG4}),
compared with results from an ab initio (DFT) simulations  of a 216-atom cell at the same temperature \cite{mazza}. 
The interatomic distance thresholds defined in Fig. \protect\ref{FIG4} are used.}
\label{TAB3}
\end{table}

\begin{table}[htbp]
\begin{center}
\begin{tabular}{c|c|c}
\hline
\hline
 & NN & DFT \\
\hline
D$_{Ge}$ (10$^{-5}$  cm$^2$/s) &  4.96 & 4.65  \\
\hline
D$_{Te}$ (10$^{-5}$  cm$^2$/s) &  3.62 & 3.93  \\
\hline
\hline
\end{tabular}
\end{center}
\caption{Diffusion coefficient of Ge and Te atoms in the 4096-atom model of liquid GeTe at 1000 K. NN results were obtained from 
the  slope of the mean square displacement versus time. The same values within two decimals are obtained from the 
integral of the velocity-velocity autocorrelation function. DFT data of a 216-atom cell at the same temperature are 
from Ref.\citenum{akola}.}
\label{TAB4}
\end{table}

\subsection{\label{amorphous} Amorphous phase}

The structural properties of a-GeTe and a-GST have been elucidated recently by ab initio simulations~\cite{akola,mazza,cara07,cara09,sosso2011}. 
In these systems Ge and Te atoms are mostly four-coordinated
and three-coordinated, respectively. Te atoms are in a defective octahedral-like environment, which resembles
the local environment of the corresponding crystalline phases. The majority of Ge atoms are in a defective octahedral
environment too, but about one quarter of Ge atoms are in a tetrahedral-like geometry.
The presence of homopolar Ge-Ge (and, in the case of GST, Ge-Sb) bonds favors the tetrahedral coordination. 

\begin{figure}[htbp!]
\centerline{\includegraphics[width=0.7\columnwidth]{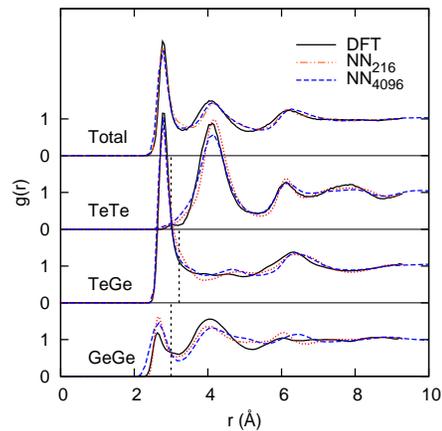}}
\caption{Total and partial pair correlation functions of amorphous GeTe from a NN molecular dynamics simulation at 300 K with a 
4096-atom and a 216-atom cell, compared with results from an ab initio simulation at the same temperature using a 216-atom 
cell \cite{mazza}. The vertical lines are the interatomic distance thresholds used to define the coordination 
numbers, 3.0 \AA\, 3.22 \AA\, and 3.0 \AA\ for Ge-Ge, Ge-Te and Te-Te bonds, respectively.}
\label{FIG6}
\end{figure}

\begin{figure}[htbp!]
\centerline{\includegraphics[width=0.5\columnwidth]{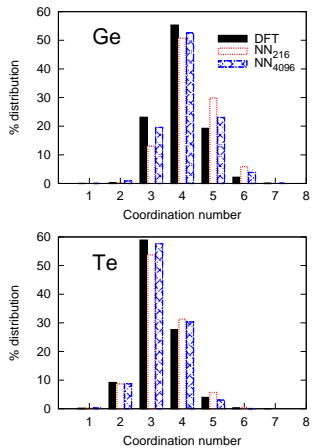}}
\caption{Distribution of coordination numbers of Ge and Te atoms in a-GeTe from NN and DFT \cite{mazza} simulations 
with 4096- and 216-atom cells. The interatomic distance thresholds defined in Fig. \ref{FIG6} are used.}
\label{FIG6bis}
\end{figure}

\begin{figure}[htbp!]
\centerline{\includegraphics[width=0.5\columnwidth]{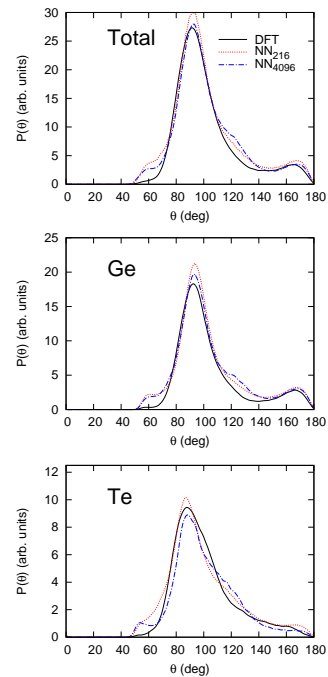}}
\caption{Total and partial angle distribution functions of amorphous GeTe from a NN molecular dynamics simulation 
at 300 K with a 4096-atom and a 216-atom cell, compared with results from an ab initio simulation at the same temperature and 
a 216-atom cell \cite{mazza}.
Partial distributions refer to X-Ge-Y and X-Te-Y triplets (X,Y= Ge or Te).}
\label{FIG7}
\end{figure}

\begin{figure*}[htbp!]
\centerline{\includegraphics[width=1.6\columnwidth]{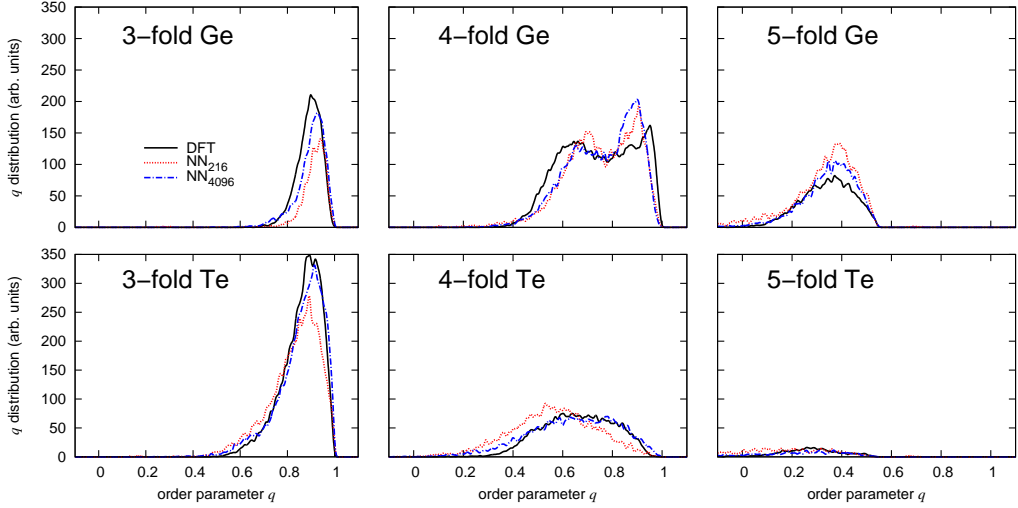}}
\caption{Order parameter $q$ for tetrahedricity (see text) for 3-, 4-, and 5-fold coordinated Ge and Te atoms in amorphous GeTe from  
NN molecular dynamics simulations at 300 K with a 4096-atom and a 216-atom cell,
compared with results from an ab initio simulation at the same temperature employing a 216-atom cell~\cite{mazza}.}
\label{FIG8}
\end{figure*}

In the
following, we compare the structural properties of models of amorphous GeTe generated by NN and ab initio simulations.
The NN amorphous phase was generated by quenching the molten sample from 1150 K to room temperature in 100 ps.
Average properties are obtained from a NVE simulation 40 ps long at an average temperature of 300 K.
Doubling or even tripling the quenching time (up to 300 ps) does not introduce sizable changes in the structural and  
vibrational properties of our NN model of amorphous GeTe. Structural properties are described in Figs.~\ref{FIG6}-\ref{FIG8}.
The partial pair correlation functions of our NN models are compared with ab initio data in Fig.~\ref{FIG6}.
By decreasing the system size from 4096-atom to 1728-atom one obtains essentially the same results.
By using instead a small 216-atom cell fluctuations in properties related to Ge-Ge homopolar bonds are found by looking at ten
different independent models. The quantities averaged over ten models are reported hereafter and are very close to those of the larger
models. The fluctuations among the different 216-atom models can be appreciated in Figs. S1-S2-S3 of the additional
material (EPAPS~\cite{epaps}).

The distribution of coordination numbers and their average values are reported in Fig.~\ref{FIG6bis} and Table~\ref{TAB5} for NN 
and ab initio~\cite{mazza} simulations.  Bond angles distribution functions  are reported in Fig.~\ref{FIG7}.

\begin{table}[htbp]
\begin{center}
\resizebox{8cm}{!} {
\begin{tabular}{c|c|c|c|c|c|c|c|c|c}
\hline
\hline
 & \multicolumn{3}{c|}{With Ge} & \multicolumn{3}{|c|}{With Te} & \multicolumn{3}{|c}{Total} \\
\hline
 & DFT & NN$_{216}$ & NN$_{4096}$ & DFT & NN$_{216}$ & NN$_{4096}$ & DFT & NN$_{216}$ & NN$_{4096}$ \\
\hline
Ge & 0.76 & 0.78 & 0.88 & 3.24 & 3.31 & 3.22 & 4.00 & 4.09 & 4.10 \\
\hline
Te & 3.24 & 3.31 & 3.22 & 0.02 & 0.04 & 0.05 & 3.27 & 3.35 & 3.27 \\
\hline
\hline
\end{tabular}}
\end{center}
\caption{Average coordination number for different pairs of atoms computed from the partial pair correlation functions of 
amorphous GeTe from  NN molecular dynamics simulations at 300 K with a 4096-atom and a 216-atom cell,
compared with results from  DFT simulations at the same temperature and a 216-atom cell \cite{mazza}.
The interatomic distance thresholds defined in Fig. \ref{FIG6} are used.}
\label{TAB5}
\end{table}

The agreement between NN and DFT data is overall very good. The largest discrepancy is on the height of the first peak of the Ge-Ge
pair correlation function. However, this misfit is partially due to fluctuations in the number of Ge-Ge bonds in the still small
216-atom cell (cf. Figs. S1-S2 in EPAPS~\cite{epaps}). In fact, by generating ten different 216-atom models with the NN potential
we observed sizable fluctuations in the average Ge-Ge coordination number ranging from 0.6 to 0.9.
A more compelling comparison would require the availability of several independent ab initio models
of a-GeTe.

\begin{figure}[htbp!]
\centerline{\includegraphics[width=0.6\columnwidth]{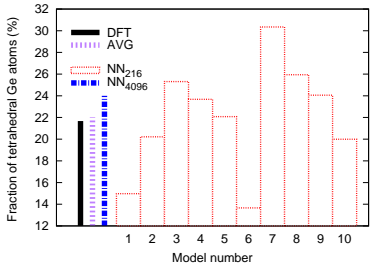}}
\caption{Fraction of tetrahedral Ge atoms in amorphous GeTe from ten statistically independent
216-atom models generated with the NN potential, compared with results from a single ab initio 216-atom model~\cite{mazza}.
The average (AVG) over the ten NN models and from the 4096-atom model are also reported.}
\label{FIG9}
\end{figure}

Another discrepancy with the DFT results is the presence of a small peak at around 60$^{\circ}$ in the angle distribution function due to a
very small fraction of three membered rings (see below). Following our previous works \cite{cara07}, we quantified the fraction of Ge atoms
in a tetrahedral geometry by computing the local order parameter $q= 1-\frac{3}{8}\sum_{i > k} (\frac{1}{3} +
cos \theta_{ijk})^2$   where the sum runs over the pairs of atoms bonded to a central atom $j$.
$q=1$ for the ideal tetrahedral geometry, $q=0$ for the six-coordinated octahedral site, and $q=5/8$ for
a four-coordinated defective octahedral site. The distribution of the local order parameter $q$ for Ge atoms
is reported in Fig.~\ref{FIG8} for  different coordination numbers. The $q$ distribution for 4-coordinated
Ge is bimodal with  peaks corresponding to  defective octahedra and tetrahedra. In contrast, the $q$-distribution 
for Te does not show any signature of the tetrahedral geometry (cf. Fig.~\ref{FIG8}).
We estimated the fraction of tetrahedral Ge atoms by integrating the $q$-distribution of 4-coordinated Ge from
0.8 to 1. This procedure was demonstrated to provide reliable values for the fraction of tetrahedral Ge from the analysis
of the Wannier functions that allow a direct identification of the tetrahedral geometry in terms of the electronic structure~\cite{sosso2011,spreafico}.
In fact, Ge in tetrahedral sites has four bonding sp$^3$-like Wannier functions, while Ge in defective octahedra has three
p-like bonding Wannier functions and one s-like lone pair. The fraction of tetrahedral Ge atoms for the ten 216-atom NN models,
for the large 4096-atom model and for the ab initio 216-atom model of Ref.~\citenum{mazza} is given in Fig.~\ref{FIG9}.
By chance the ab initio value is very close to the value obtained by averaging over the ten
216-atom NN models while the fraction of tetrahedral Ge in the large 4096-atom cell is 24~$\%$, 
a value very close to the average over the ten small cells and also to the result obtained for a 1728-atom model (22~$\%$). 
The concentration of tetrahedra depends on the fraction of Ge-Ge homopolar bonds, 
which is the property more affected by finite size effects in the small 216-atom cell.

\begin{figure}[htbp!]
\centerline{\includegraphics[width=0.6\columnwidth]{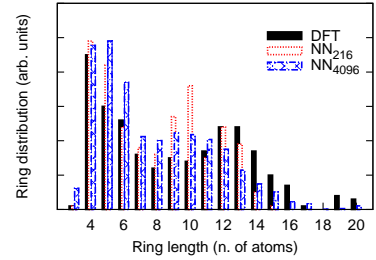}}
\caption{Distribution of ring lengths in the 4096- and 216-atom NN models and in the DFT 216-atom model of Ref.~\citenum{mazza}.}
\label{FIG10}
\end{figure}

\begin{figure}[htbp!]
\centerline{\includegraphics[width=0.6\columnwidth]{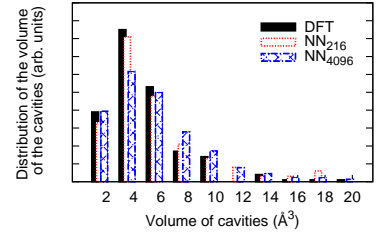}}
\caption{Distribution of the volume of cavities in the 4096- and 216-atom NN models and in the DFT 216-atom 
model of Refs.~\citenum{mazza,caraSbTe}.} 
\label{FIG10bis}
\end{figure}

Turning now to the medium range order, we report in Fig.~\ref{FIG10} the distribution of ring lengths computed  according to 
Ref.~\citenum{franz} for the large and small NN models and for the small DFT model of Ref.~\citenum{mazza}. The amorphous phases of 
GeTe and GST have been shown to display a large concentration of nanocavities~\cite{akola}. The distribution of the volume of 
nanocavities computed according to the definition of Ref.~\citenum{alin} and the algorithm of Ref.~\citenum{medvedev,VNP} is 
compared in Fig.~\ref{FIG10bis} for the NN and DFT models.
The same scheme for the calculation of nanocavities was applied in our previous works on different phase change materials~\cite{caraSbTe}.
These comparisons show that the  agreement between NN and DFT results is very good  for  the medium range order as well.

We also optimized the density of the amorphous model at zero temperature by interpolating the energy-volume
points with a Murnaghan equation of state. We obtained an equilibrium density of 0.03351 atoms \AA$^{-3}$ to be 
compared with the value of 0.03156 atoms \AA$^{-3}$ resulting from the ab initio equation of state of a
216-atom cell with the BZ integration restricted to the $\Gamma$-point~\cite{mazza}. The experimental equilibrium
density~\cite{expd} of a-GeTe is 0.03327 atoms~\AA$^{-3}$. The NN and ab initio bulk moduli of a-GeTe are 17~GPa and 14~GPa, respectively.

\begin{figure}[htbp!]
\centerline{\includegraphics[width=0.6\columnwidth,angle=-90]{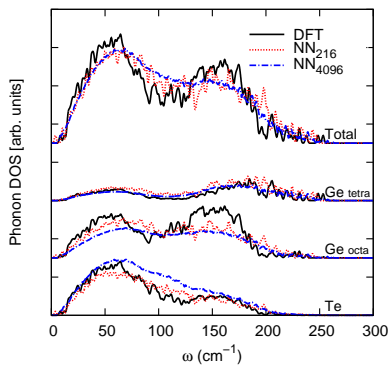}}
\caption{Phonon density of states of amorphous GeTe from 4096- and 216-atom NN models. Results for a single 216-atom DFT model \cite{mazza}
are also reported. Only phonons at the supercell $\Gamma$ point are
considered. The higher smoothness of 4096-atom NN data is due to the larger supercell used. Projections of the DOS on
the different atomic species (Te atoms and Ge atoms in tetrahedral and defective octahedral geometries) are also shown.}
\label{FIG11}
\end{figure}

Concerning the vibrational properties, the phonon density of states of amorphous GeTe from NN and ab initio simulations are 
compared in Fig.~\ref{FIG11}. Phonon frequencies are computed by diagonalizing the dynamical matrix obtained in turn from the 
variation of atomic forces due to finite atomic displacements 0.02~\AA\ large. Only phonons with the periodicity of our 
supercells ($\Gamma$-point phonons) were considered.
Ab initio phonons are computed in a 216-atom cell~\cite{mazza} while NN phonons are obtained from the 4096-atom and 216-atom cells.
Projections on the different type of atoms (Te, Ge in tetrahedral and defective octahedral geometries) are also shown.

In an amorphous material, phonons display
localization properties, which depend on frequency. To address this
issue and following our previous ab initio works~\cite{mazza}, we computed the inverse participation ratio ($IPR$) of
the $j$-th vibrational mode  defined as

\begin{equation}
\label{Eq_ipr}
 IPR=\frac{\sum_{\kappa}\left| \frac{{\bf e}(j,\kappa)}{\sqrt{M_{\kappa}}}\right|^4}
{\left(\sum_{\kappa}\frac{\left| {\bf
e}(j,\kappa)\right|^2}{M_{\kappa}}\right)^2} .
\end{equation}

\begin{figure}[htbp!]
\vspace{0.5cm}
\centerline{\includegraphics[width=0.6\columnwidth,angle=-90]{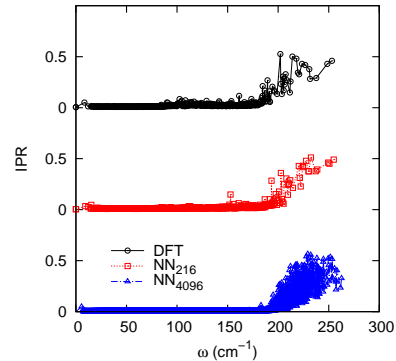}}
\caption{Inverse participation ratio of phonons in a-GeTe  from 4096- and 216-atom NN models.  Results for a single 216-atom DFT model
\cite{mazza} are also reported.}
\label{FIG12}
\end{figure}

Here ${\bf e}(j,\kappa)$ are phonon eigenvectors and the sum over
$\kappa$ runs over the $N$ atoms in the unit cell with masses
$M_{\kappa}$. According to this definition, the value of $IPR$
varies from $1/N$ for a completely delocalized phonon, to one for
a mode completely localized on a single atom. The values of $IPR$
for the NN and ab initio models of a-GeTe are reported in Fig.~\ref{FIG12}.
The NN potential reproduces the strong localization on tetrahedra of phonons above 200~cm$^{-1}$.
The overall shape and frequency range of the phonon DOS is reasonably reproduced by the NN potential. A discrepancy
is present in the relative height of the two main structures at 50~cm$^{-1}$ and 150~cm$^{-1}$, which, however, might be partially due to
the still small size of the 216-atom cell. Actually the DOS of the different 216-atom models generated by the NN potentials are all very
similar, but  somehow different from the large 4096-atom cell due to a size effect.

\section{\label{conclusions} CONCLUSIONS}

In summary, a NN potential for the phase change material GeTe has been created and
tested to reproduce the properties of crystalline, liquid and amorphous phases.
The NN potential has been validated by comparing the results on structural and dynamical
properties of the bulk phases of GeTe with our previous data from DFT calculations~\cite{mazza}.
The development of a classical potential with close to ab initio accuracy represents a
breakthrough in the simulation of phase change materials, as it will allow  addressing several
key issues on the properties of this class of materials that are presently  beyond the reach of ab initio simulations.
The study of thermal conductivity in the amorphous phase~\cite{thermal} and the dynamics of homogeneous and heterogeneous
crystallization  are few examples of useful follow-on developments of this work, which promises to improve our microscopic
understanding of the operation of phase change memories.

\begin{acknowledgments}
We thankfully acknowledge the computational resources by DEISA Consortium under projects NETPHASE,
by CSCS (Manno, Switzerland) and by the ISCRA Initiative at Cineca.
This work has been partially supported by Regione Lombardia and CILEA Consortium
through a LISA Initiative (Laboratory for Interdisciplinary Advanced
Simulation) 2011 grant [link: http://lisa.cilea.it ],
by the Cariplo Foundation through project Monads and by MURST through the program Prin08.
JB thanks the DFG for financial support (Emmy Noether program).
\end{acknowledgments}

\newpage
\thispagestyle{empty}
\mbox{}

\begin{figure}[b]
\begin{minipage}[b]{1.8\columnwidth}
{\bf {\large Additional materials}}
\renewcommand\figurename{Fig.~S$\!\!$}
\centerline{\includegraphics[width=0.5\columnwidth]{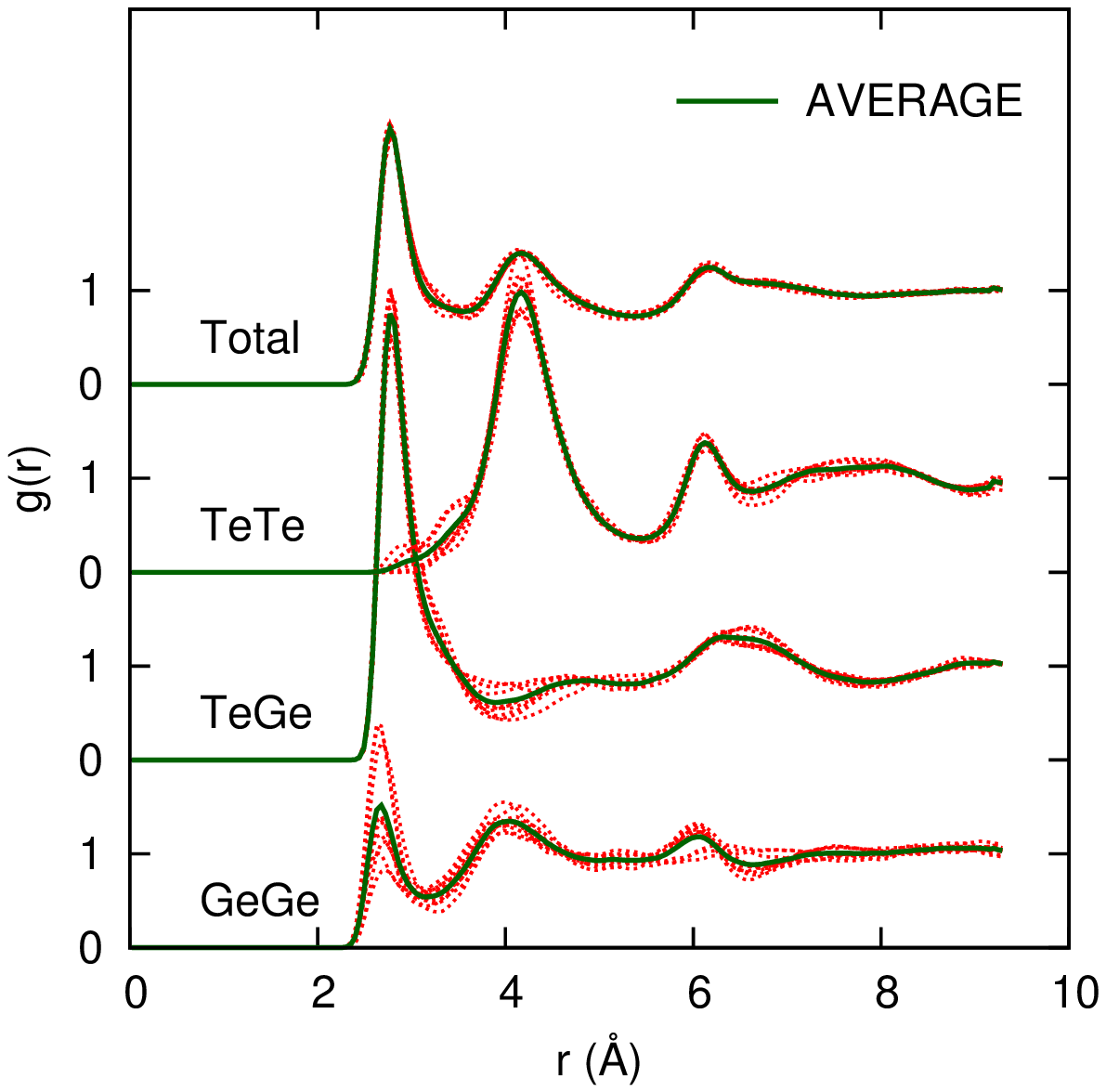}}
\caption{Total and partial pair correlation functions of amorphous GeTe from NN molecular dynamics simulations at 300 K of ten
statistically indipendent 216-atom models. Average values are also reported.}
\label{Fig_fgdr}
\end{minipage}
\hspace{0.5cm}
\begin{minipage}[b]{1.8\columnwidth}
\renewcommand\figurename{Fig.~S$\!\!$}
\centerline{\includegraphics[width=0.5\columnwidth]{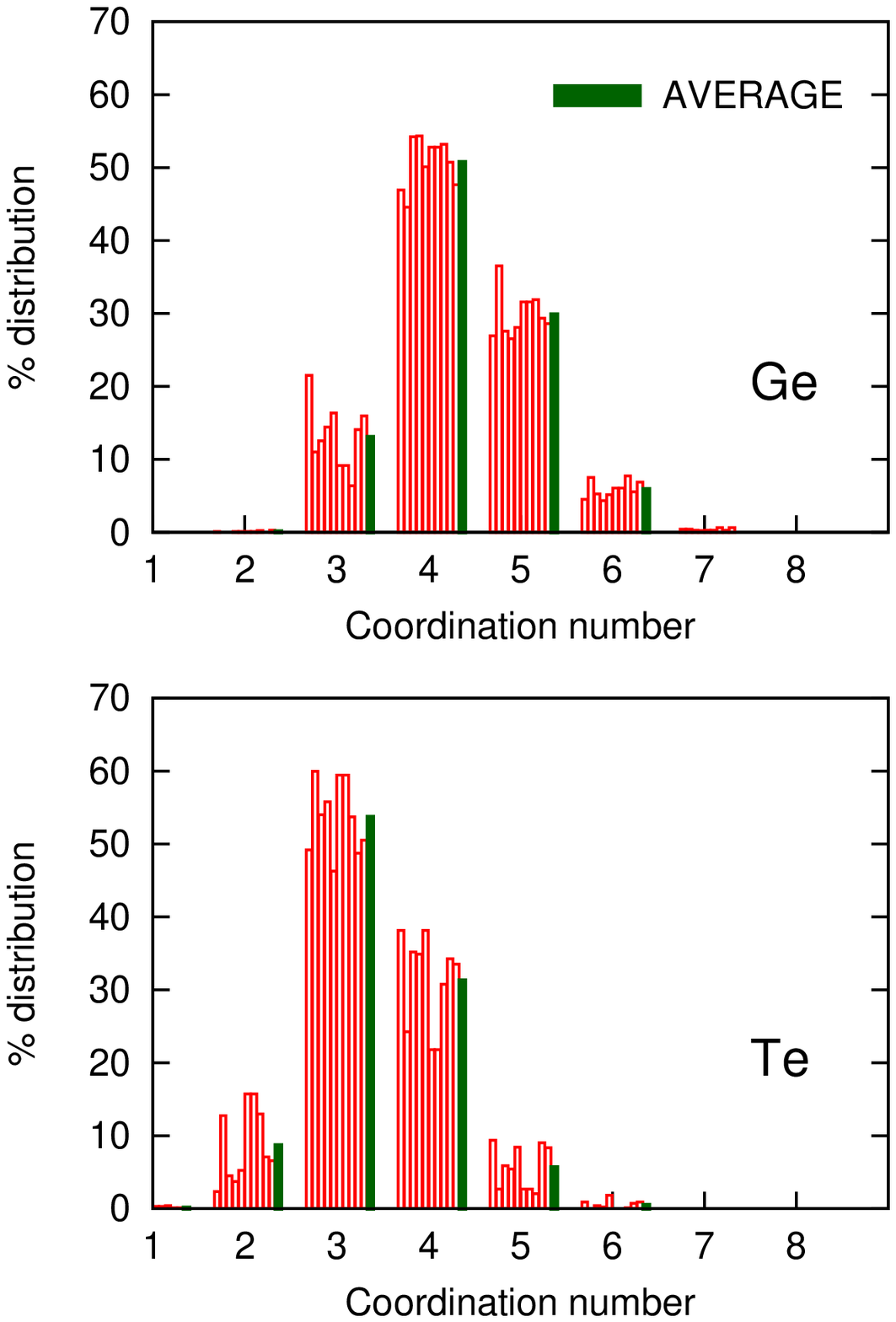}}
\caption{Distribution of coordination numbers of Ge and Te atoms of amorphous GeTe from NN molecular dynamics simulations at 300 K of ten
statistically indipendent 216-atom models. Average values are also reported.}
\label{Fig_fcn}
\end{minipage}
\end{figure}

\begin{figure}[b]
\begin{minipage}[b]{1.8\columnwidth}
\renewcommand\figurename{Fig.~S$\!\!$}
\centerline{\includegraphics[width=0.5\columnwidth]{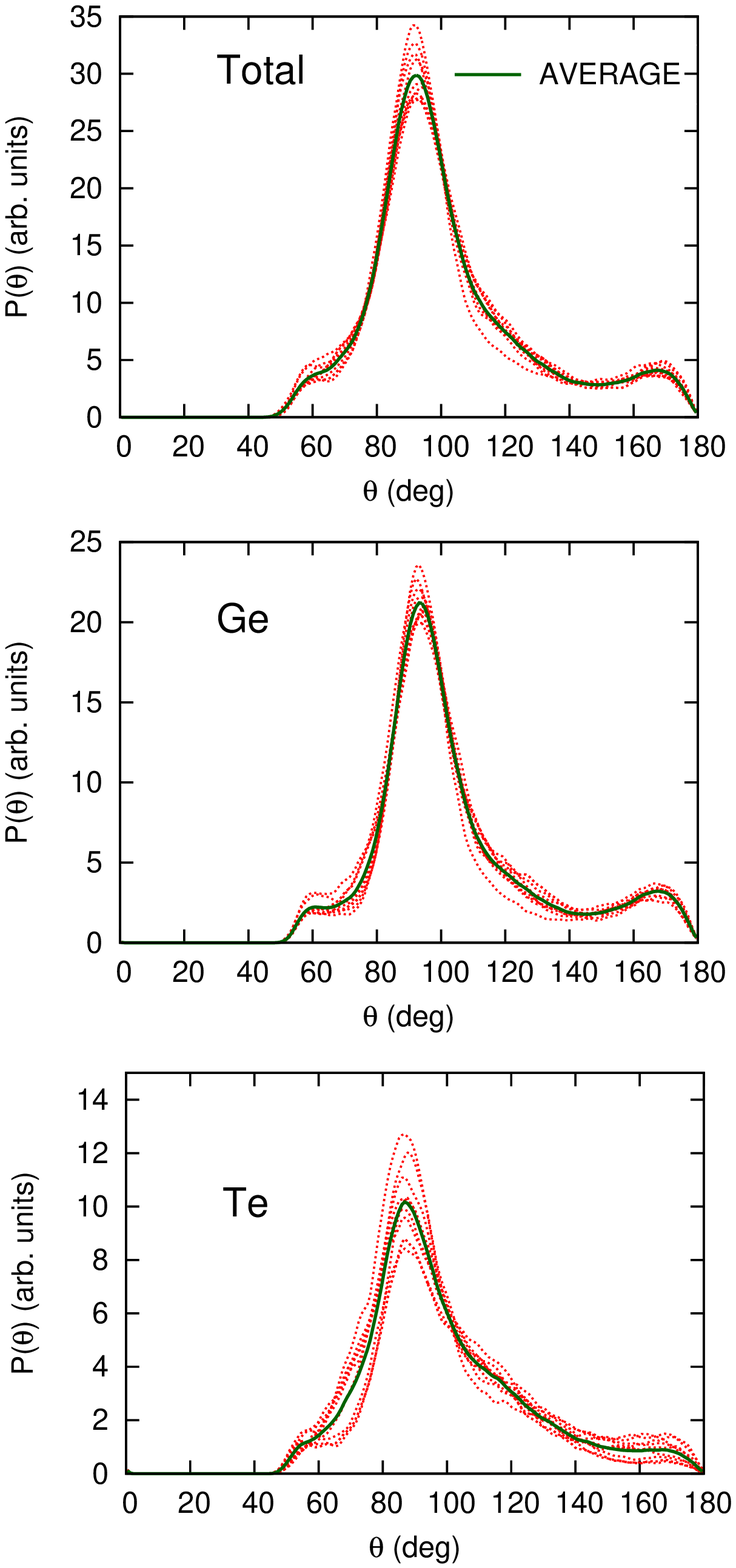}}
\caption{Total and partial angle distribution functions of amorphous GeTe from NN molecular dynamics simulations at 300 K of ten
statistically indipendent 216-atom models. Average values are also reported. Partial distributions refer to X-Ge-Y
and X-Te-Y triplets (X,Y= Ge or Te).}
\label{Fig_fpt}
\end{minipage}
\end{figure}


\begin{thebibliography}{99}

\bibitem{1} M. Wuttig and N. Yamada, Nat. Mater. {\bf 6}, 824 (2007).
\bibitem{2} A. Pirovano, A. L. Lacaita, A. Benvenuti, F. Pellizzer, and R. Bez, IEEE Trans. Electron. Dev. {\bf 51}, 452 (2004).
\bibitem{3} A.L. Lacaita and D.J. Wouters, Phys. Stat. Sol. A {\bf 205}, 2281 (2008).
\bibitem{lencerReview} D. Lencer, M. Salinga, and M. Wuttig, Adv. Mat. {\bf 23}, 2030 (2011).
\bibitem{welrev} S. Raoux, W. Welnic, and D. Ielmini, Chem. Rev. {\bf 110}, 240 (2010).
\bibitem{cara07} S. Caravati, M. Bernasconi, T. D. K\"{u}hne, M. Krack, and M. Parrinello, Appl. Phys. Lett. {\bf 91}, 171906 (2007).
\bibitem{cara09} S. Caravati, M. Bernasconi, T. D. K\"{u}hne, M. Krack, and M. Parrinello, J. Phys. Cond. Mat. {\bf 21}, 255501 (2009);
errata {\bf 21}, 499803 (2009); {\bf 22}, 399801 (2010).
\bibitem{akola} J. Akola, and R. O. Jones, Phys. Rev. B {\bf 76}, 235201 (2007).
\bibitem{7} J. Behler, and M. Parrinello, Phys. Rev. Lett. {\bf 98}, 146401 (2007).
\bibitem{8} J. Behler, R. Marto\v{n}\'{a}k, D. Donadio, and M. Parrinello, Phys. Rev. Lett. {\bf 100}, 185501 (2008).
\bibitem{pssbbehler} J. Behler, R. Marto\v{n}\'{a}k, D. Donadio, and M. Parrinello, Phys. Stat. Sol. B {\bf 245}, 2618 (2008).
\bibitem{9} R.Z. Khaliullin, H. Eshet, T.D. K\"uhne, J. Behler, and M. Parrinello, Phys. Rev. B {\bf 81}, 10010 (2010).
\bibitem{khaliullin2011}  R.~Z. Khaliullin, H. Eshet, T.~D. K\"uhne, J. Behler, and M. Parrinello, Nat. Mater. {\bf 10}, 693 (2011).
\bibitem{10} H. Eshet, R.Z. Khaliullin, T.D. K\"uhne, J. Behler, and M. Parrinello, Phys. Rev. B {\bf 81}, 184107 (2010).
\bibitem{11} N. Artrith, T. Morawietz, and J. Behler, Phys. Rev. B {\bf 83}, 153101 (2011).
\bibitem{artrith} N. Artrith, and J. Behler, submitted (2011).
\bibitem{mazza} R. Mazzarello, S. Caravati, S. Angioletti-Uberti, M. Bernasconi, and M. Parrinello, Phys. Rev. Lett. {\bf 104}, 085503 (2010); erratum {\bf 107}, 039902 (2011).
\bibitem{bishop} C.M. Bishop, ``Neural Networks for Pattern Recognition'', Oxford University Press, 1995.
\bibitem{handely} C.~M. Handley, and P.~L.~A. Popelier, J. Phys. Chem. A {\bf 114}, 3371 (2010). 
\bibitem{pccpbehler} J. Behler, Phys. Chem. Chem. Phys. {\bf 13}, 17930 (2011).
\bibitem{hornik} K. Hornik, M. Stinchcombe, and H. White, Neural Networks {\bf 2}, 359 (1989).
\bibitem{cybenko} G. Cybenko, Math. Contr. Sign. Systems {\bf 2}, 303 (1989).
\bibitem{gassner} H. Gassner, M. Probst, A. Lauenstein, and K. Hermansson, J. Phys. Chem. A {\bf 102}, 4596 (1998).
\bibitem{p0420} D.~F.~R. Brown, M.~N. Gibbs, and D.~C. Clary, J. Chem. Phys. {\bf 105}, 7597 (1996).
\bibitem{p0831} L.~M. Raff, M. Malshe, M. Hagan, D.~I. Doughan, M.~G. Rockley, and R. Komanduri, J. Chem. Phys. {\bf 122}, 084104 (2005).
\bibitem{p0826} S. Manzhos, X. Wang, R. Dawes, and T. Carrington, Jr., J. Phys. Chem. A {\bf 110}, 5295 (2006).
\bibitem{p2391} S. Houlding, S.~Y. Liem, P.~L.~A. Popelier, Int. J. Quantum Chem. {\bf 107}, 2817 (2007).
\bibitem{blank} T.~B. Blank, S.~D. Brown, A.~W. Calhoun, and D.~J. Doren, J. Chem. Phys. {\bf 103}, 4129 (1995).
\bibitem{lorenz03} S. Lorenz, A. Gro\ss, and M. Scheffler, Chem. Phys. Lett. {\bf 395}, 210 (2004).
\bibitem{p1786} J. Ludwig, and D.G. Vlachos, J. Chem. Phys. {\bf 127}, 154716 (2007).
\bibitem{behler2007} J. Behler, S. Lorenz, and K. Reuter, J. Chem. Phys. {\bf 127}, 014705 (2007).
\bibitem{behler2008} J. Behler, K. Reuter, and M. Scheffler, Phys. Rev. B {\bf 77}, 115421 (2008).
\bibitem{carbogno} C. Carbogno, J. Behler, A. Gro\ss, and K. Reuter, Phys. Rev. Lett. {\bf 101}, 096104 (2008). 
\bibitem{17} J. Behler, J. Chem. Phys. {\bf 134}, 074106 (2011).
\bibitem{morawietz} T. Morawietz, V. Sharma, and J. Behler, submitted (2011).
\bibitem{15} R. Marto\v{n}\'{a}k, A. Laio, and M. Parrinello, Phys. Rev. Lett. {\bf 90}, 075503 (2003).
\bibitem{16} J. VandeVondele, M. Krack, F. Mohamed, M. Parrinello, T. Chassaing, and J. Hutter, Comp. Phys. Comm. {\bf 167}, 103 (2005); 
M. Krack, and M. Parrinello, {\it High Performance Computing
 in Chemistry}, edited by J. Grotendorst (NIC, Julich, 2004),
Vol. {\bf 25}, pp. 29-51; http://cp2k.berlios.de.
\bibitem{MP} H. J. Monkhorst, and J. D. Pack, Phys. Rev. B {\bf 13}, 5188 (1976).
\bibitem{12} P. Giannozzi et al., J. Phys. Cond. Mat. {\bf 21}, 395502 (2009).
\bibitem{13} J. P. Perdew, K. Burke, and M. Ernzerhof, Phys. Rev. Lett. {\bf 77}, 3865 (1996).
\bibitem{21} RuNNer: A Neural Network Code for High-Dimensional Potential-Energy Surfaces, J\"org Behler, Lehrstuhl f\"ur Theoretische
Chemie, Ruhr-Universit\"at Bochum, Germany.
\bibitem{22} J.~W. Ponder, Department of Chemistry, Washington University, 
Saint Louis, USA, http://dasher.wustl.edu/tinker/
\bibitem{bere} H.J.C. Berendsen, J.P.M. Postma, W.F. van Gunsteren, A. DiNola, and J. R. Haak, J. Chem. Phys. {\bf 81}, 3684 (1984).
\bibitem{goldak} J. Goldak, C.S. Barrett, D. Innes, and W. Youdelis, J. Chem. Phys. {\bf 44}, 3323 (1966).
\bibitem{murna} F.D. Murnaghan, PNAS {\bf 30}, 244 (1944).
\bibitem{lencer} D. Lencer, M. Salinga, B. Grabowski, T. Hickel, J. Neugebauer, and M. Wuttig, Nat. Mater. {\bf 7 }, 972 (2008).
\bibitem{ratyPhon} R. Shaltaf, E. Durgun, J.Y. Raty, P. Ghosez, and X. Gonze, Phys. Rev. B {\bf 78}, 205203 (2008).
\bibitem{adens} G.E. Ghezzi, J.Y. Raty, S. Maitrejean, A. Roule, E. Elkaim, and F. Hippert, Appl. Phys. Lett {\bf 99}, 151906 (2011).
\bibitem{caraSbTe} S. Caravati, M. Bernasconi, and M. Parrinello, Phys. Rev. B {\bf 81}, 014201 (2010).
\bibitem{spreafico} E. Spreafico, S. Caravati, and M. Bernasconi, Phys. Rev. B {\bf 83}, 144205 (2011).
\bibitem{sosso2011} G.C. Sosso, S. Caravati, R.Mazzarello, and M. Bernasconi, Phys. Rev. B {\bf 83}, 134201 (2011).
\bibitem{epaps} Additional materials can be found at the very bottom of this document.
\bibitem{franz} D. S. Franzblau, Phys. Rev. B {\bf 44}, 4925 (1991).
\bibitem{alin} M. G. Alinchenko, A. V. Anikeenko, N. N. Medvedev, V. P. Voloshin, M. Mezei, and P. Jedlovszky, J. Phys. Chem. B {\bf 108},
19056 (2004).
\bibitem{medvedev} N.N. Medvedev, V.P. Voloshin, V.A. Luchnikov, and M.L. Gavrilova, J. Comp. Chem. {\bf 27}, 1676 (2006).
\bibitem{VNP} http://www.kinetics.nsc.ru/sms/?Software:VNP
\bibitem{expd} G.E. Ghezzi, J.Y. Raty, S. Maitrejean, A. Roule, E. Elkaim, and F. Hippert, Appl. Phys. Lett. {\bf 99}, 151906 (2011).
\bibitem{thermal} G.C. Sosso, D. Donadio, S. Caravati, J. Behler, and M. Bernasconi, submitted (2012).

\end{thebibliography}
\end{document}